\documentclass[runningheads]{llncs}
\usepackage{times}
\usepackage{xcolor}
\usepackage{soul}
\usepackage[utf8]{inputenc}

\usepackage{comment}
\usepackage{color}
\usepackage{framed}
\usepackage[noadjust]{cite}

\usepackage{epsfig}
\usepackage{graphicx}

\usepackage[ruled,vlined,linesnumbered]{algorithm2e}
\usepackage{url}
\usepackage{multirow}
\usepackage{enumitem}
\usepackage{moresize}
\usepackage{epstopdf}
\usepackage{array}

\usepackage{diagbox}
\usepackage[export]{adjustbox}

\usepackage{tcolorbox}

\usepackage{colortbl}

\usepackage{multirow} 
\usepackage{extarrows}

\makeatletter
\newcommand{\thickhline}{%
    \noalign {\ifnum 0=`}\fi \hrule height 1pt
    \futurelet \reserved@a \@xhline
}
\newcolumntype{"}{@{\hskip\tabcolsep\vrule width 1pt\hskip\tabcolsep}}
\makeatother

\newcommand{\tool}{{\textit{DeepHunter}\xspace}}
\newcommand{\affinetransform}{\mathcal{G}}
\newcommand{\pixeltransform}{\mathcal{P}}
\newcommand{\image}{\mathcal{I}}

\usepackage{wrapfig}
\usepackage{subfigure}
\usepackage[hang,flushmargin]{footmisc} % remove footnote indentation

\usepackage[font=footnotesize,labelfont=bf]{caption}

\usepackage{xspace}

\makeatletter
\DeclareRobustCommand\onedot{\futurelet\@let@token\@onedot}
\def\@onedot{\ifx\@let@token.\else.\null\fi\xspace}
\def\eg{\emph{e.g}\onedot} 
\def\ie{\emph{i.e}\onedot} 
\def\cf{\emph{c.f}\onedot} 
\def\etc{\emph{etc}\onedot} 
 
\def\etal{\emph{et al}\onedot}
\makeatother

\begin{document}

\clubpenalty = 10000
\widowpenalty = 10000
\displaywidowpenalty = 10000

\title{{\tool}: Hunting Deep Neural Network Defects via Coverage-Guided Fuzzing}

\author{Xiaofei Xie\inst{1}\and
Lei Ma\inst{2} \and
Felix Juefei-Xu\inst{3}\and
Hongxu Chen\inst{1}\and
Minhui Xue\inst{1} \and \\ 
Bo Li\inst{4}\and
Yang Liu\inst{1}\and
Jianjun Zhao\inst{5}\and
Jianxiong Yin\inst{6} \and 
Simon See\inst{6}}

\authorrunning{X. Xie et al.}

\institute{Nanyang Technological University \and Harbin Institute of Technology \and
Carnegie Mellon University \and  University of Illinois at Urbana–Champaign \and Kyushu University \and
NVIDIA AI Technology Center
}

\maketitle

\begin{abstract}
In company with the data explosion over the past decade, deep neural network (DNN) based software has experienced unprecedented leap and is becoming the key driving force of many novel industrial applications, including many safety-critical scenarios such as autonomous driving. Despite great success achieved in various human intelligence tasks, similar to traditional software, DNNs could also exhibit incorrect behaviors caused by hidden defects causing severe accidents and losses.
In this paper, we propose {\tool}, an automated fuzz testing framework for hunting potential defects of general-purpose DNNs. {\tool} performs metamorphic mutation to generate new semantically preserved tests, and leverages multiple plugable coverage criteria as feedback to guide the test generation from different perspectives. To be scalable towards practical-sized DNNs, {\tool} maintains multiple tests in a batch, and prioritizes the tests selection based on active feedback. The effectiveness of {\tool} is extensively investigated on 3 popular datasets (MNIST, CIFAR-10, ImageNet) and 7 DNNs with diverse complexities, under large set of 6 coverage criteria as feedback. The large-scale experiments demonstrate that {\tool} can (1) significantly boost the coverage with guidance; 
(2) generate useful tests to detect erroneous behaviors and facilitate the DNN model quality evaluation; (3) accurately capture potential defects during DNN quantization for platform migration.


\end{abstract}

\section{Introduction}
Recently great success has been achieved by artificial intelligence (AI) systems, such as IBM's Watson \cite{kelly2013smart}, Amazon Alexa \cite{Amazon_Alexa}, as well as DeepMind's Atari \cite{mnih2015human} and AlphaGo~\cite{alphago}. We are now engaged in new AI development and deployment at an unprecedented speed and scale. Deep learning (DL), or deep neural network (DNN) systems, is becoming the paramount ingredient of various kinds of AI-enabled applications, such as speech processing \cite{hinton2012deep}, medical diagnostics \cite{ciresan2012deep}, image processing \cite{ciregan2012multi}, and robotics~\cite{zhang2015towards}, across various implementation platforms such as TensorFlow \cite{tensorflow}, Keras~\cite{keras}, PyTorch~\cite{paszke2017automatic}.
While DNNs are permeating all industry verticals, it has brought to our attention that DNNs as software 2.0 \cite{sw2} should be more extensively tested or verified. DNNs definitely deserve more scrutiny than the current practice before they are deployed to safety- and mission-critical applications.

The quality assurance of DNN-based software is still immature, it has caused great losses, such as a Google car accident~\cite{google_crash} and an Uber car crash~\cite{uber_crash}. Systemic and effective testing frameworks for reliably detecting defects and vulnerabilities in real-world sized DNN-based software are in great demand.

In traditional software realm, coverage-guided fuzz~(CGF) testing is a well-established technique for defects and vulnerability detection. It helps detect thousands of bugs and vulnerabilities issues in modern software, many of which have been existing for decades~\cite{AFL,libFuzzer,honggfuzz,fse2018FOT, rawat2017vuzzer,ccs2016AFLFast,sp2017Skyfire}. CGF performs systematic random mutations on inputs and generates test inputs to drive the software into diverse corner-case states.
The major components of the state-of-the-art CGFs often include mutation, feedback guidance, and fuzzing strategy, among which the feedback guidance can provide valuable adjustment to the fuzzing strategy and can significantly improve the efficiency of a fuzzing algorithm. 

However, due to the fundamental difference between traditional and DNN based software, traditional fuzzing elements could not be directly applied to DNN fuzzing. For example, the mutations and the feedback are all different in many ways. For traditional software, the mutation is usually rather random and frequently generates \emph{invalid} (or meaningless) seeds that will be rejected by the sanity check in the program quite early. As a result, general-purpose fuzzers usually can only find shallow bugs, such as parsing errors and improper input validations. On the other hand, the input of DNN software typically requires special formats, and inputs that violate the format specifications will be rejected even before the learning procedure starts. Therefore, it is considered more cost-effective to customize a DNN-aware mutation strategy based on some intermediate representations rather than the raw data. 

Another challenge in DNN software is that the goal is no longer detecting the \emph{vulnerabilities} or \emph{crashes} in software; rather, we now shift our focus to the \emph{functionality} of DNN results. This is more like a differential testing scenario where we need to additionally distinguish anomaly results from acceptable differences. Furthermore, the study on DNN based software testing is still at its early stage, and whether existing techniques, in particular, \emph{coverage criteria}~\cite{pei2017deepxplore,tian2018deeptest,deepguage}, can provide meaningful guidance to DNN fuzzing still lacks extensive and in-depth investigation.

In this work, we are poised to answer these questions and highlight the following contributions:
\begin{itemize}[wide=1pt,leftmargin=10pt]
    \item{We propose a general-purpose coverage guided fuzzing framework {\tool} to systematically test DNN based software, which is among the earliest studies to perform feedback-guided testing for DNNs. The design of {\tool} takes into consideration the unique characteristics of testing DNNs and the scalability towards practical-sized DNNs. (1) In particular, the test execution could be easily paralleled. We propose a batch-based strategy and leverage it to maintain high throughput in obtaining fuzzing results. (2) To enable large-scale automated generation of new test inputs within valid domain, we propose a metamorphic mutation based test generation technique, which preserves the input semantics before and after mutation. (3) We propose to guide the test generation with plugable feedback analysis components, including a set of 6 testing criteria of different granularity, to further guide the fuzzing procedures.}
    
    \item{We have performed a large-scale empirical study to evaluate the usefulness of {\tool} in systematically generating tests for coverage enhancement, guided by the 6 recently-proposed coverage criteria.}
    
    \item{We further investigate how each of the very recently-proposed testing coverage criteria helps to guide the fuzzing for, (1) DNN model quality evaluation, (2) error-behavior detection, and (3) defect introduction of quantization for platform migration.
    }

    \end{itemize}  
Overall, we find that {\tool} can effectively generate useful tests in general, in terms of (1) improving target coverage, (2) evaluating DNN model quality; (3) detecting erroneous behaviours, as well as (4) capturing  the sensitive cases where quantized DNN version fails.
To the best of our knowledge, this work is by far among the largest scaled empirical evaluations for DNN testing, using 3 datasets~(including ImageNet), 7 DNN models (with large ones like VGG-16, ResNet-50),
and a set of 6 coverage criteria as CGF guidance. We will make {\tool} publicly available as an open framework to facilitate further comparative studies on DNN testing.

\begin{figure}
\centering
\includegraphics[width=1\columnwidth]{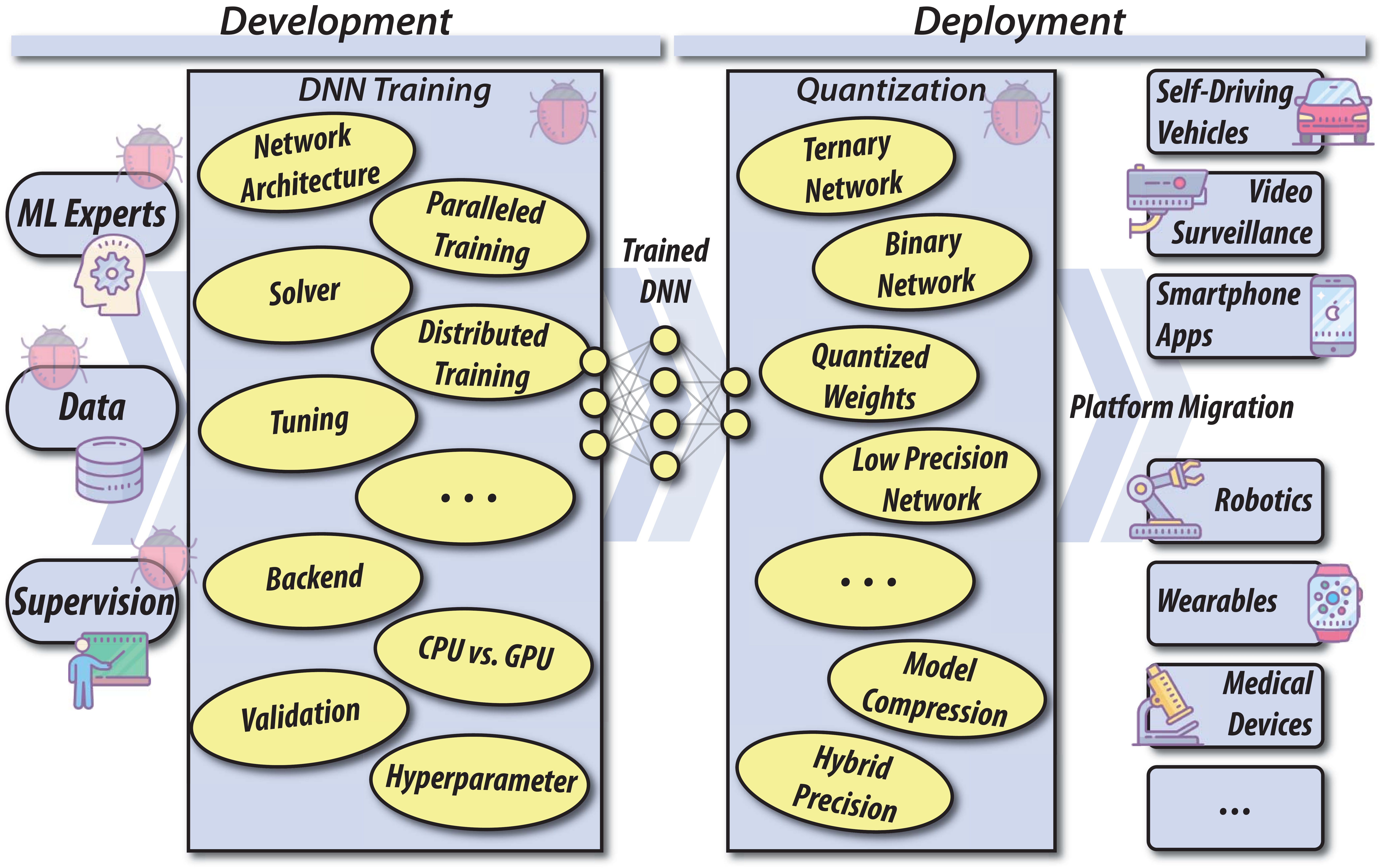}
\caption{The development and deployment process for general DNN based software. Our proposed {\tool} is dedicated to assessing the DNN software quality and hunting the defects therein.}
\label{fig:teaser}
\vspace{-3mm}
\end{figure}

\section{Preliminaries}
\label{sec:background}

\subsection{DNN Software Development and Deployment}

Over the past several decades, software development methodology \cite{Pressman:2009:SEP:1593949,Ruparelia:2010:SDL:1764810.1764814} has been well-established for traditional software, with many experiences and practices widely applied in software industry.  
A common software life cycle often consists of several key stages such as requirement analysis, design, implementation, testing, deployment, and maintenance.
These development methodology and principles also generally apply to DNN based software.
However, different from traditional software, deep learning defines a new data-driven programming paradigm, and keeps its artifact in form of an encoded deep neural network structure and neuron connection weight matrix. Such unique features bring some new challenges for quality assurance of DNN based software, especially in DNN development and deployment phases~\cite{2018arXiv181004538M}.

Figure~\ref{fig:teaser} gives an overview of the state-of-the-practice DNN software development and deployment. The development phase transforms knowledge of the machine learning experts, the prepared data, and the associated supervision signals into a deep neural network for particular tasks at hand. The training of a DNN involves many tuning knobs, such as the choices of network architecture, backend training framework, solver, and hyper-parameters. Also, one may need to consider the communication overhead, the model parallelization, the data parallelization, the CPU and GPU hybrid training, \etc.

Once an applicable DNN model is ready, it will oftentimes go through either quantization, or platform migration, or both, before being deployed to end-user applications, such as self-driving cars, video surveillance, smartphones, and wearable devices. This is because the training phase requires a vast amount of computation and energy resources. As the model size and the complexity of the tasks grow, more data are needed to train the network till reaching optimality, which could spend days, if not weeks, in training on high-performance GPU clusters. On the other hand, delete the deployment of the DNN models is usually into a resource constrained environment with limited computation, storage, and power. Therefore, when migrating from one platform to another, \eg, GPU-cluster trained DNNs to be deployed onto embedded systems or mobile devices, the DNNs usually need to go through a ``slimming'' process via quantization. 

Quantization of DNNs has been widely studied and is considered as one of the most effective approaches to meet the extreme memory and computation requirements that DNNs demand. Studies have shown that to maintain similar level of accuracy and DNN performance, full precision 32-bit floating point weights may not be necessary \cite{Jacob_2018_CVPR_integer,gupta2015deep,courbariaux2014training,wu2018training,binaryNet,binaryConnect,xnorNet,zhu2016trained,hubara2017quantized,felix_cvpr18_pnn,felix_cvpr17_lbcnn,han2015deep}. 
One can quantize the weights to much lower bits~(\eg, from 32-bit floating to 16-bit or to mixed 32 and 16 precision) in order to greatly reduce the model footprint and energy consumption, which has been commonly adopted for industrial level DNN software deployment~\cite{tensorRT}.

However, as depicted in Figure~\ref{fig:teaser}, defects might be introduced in both the development and deployment phases. For example, data collection, training program implementation, training execution, \etc, could all introduce potential defects at the DNN development stage. Similarly, quantization and platform migration can also introduce defects either due to quantization operator or compatibility issues.

Together, they are among the prime suspects for causing unexpected behaviors and vulnerabilities in DNN software products.
% --------------
The current de facto practice mainly relies on test accuracy to assess the quality of DNNs. However, this is still insufficient especially when the quality of test data is low.
A low quality test only \emph{partially} measures the DNN quality, and is unsuitable to provide insights to defects and vulnerabilities DNN software, causing some fatal defects missed without giving any feedback to the DNN developers.

\subsection{Coverage-based Grey-box Fuzzing}

Fuzzing has gained its popularity in academia and industry due to its scalability and effectivenss in generating useful tests for defect detection. Based on awareness of the target program structure, fuzzers can be classified as black-box~\cite{PeachFuzz}, white-box~\cite{cadar2008klee} or grey-box~\cite{AFL}. One of the most successful techniques is coverage-based grey-box fuzzing (CGF), which strikes a balance between effectiveness and efficiency by using code coverage as feedback. Many state-of-the-art CGFs, such as AFL~\cite{AFL}, libFuzzer~\cite{libFuzzer} and VUzzer~\cite{rawat2017vuzzer}, have been widely used and proven to be effective.

Given a target program, CGF uses a lightweight instrumentation to collect the coverage information during fuzzing.
A typical CGF usually performs the following loop~\cite{gan2018collafl}: (1) selecting seeds from the seed pool; (2) mutating the seed a certain number of times to generate new tests with mutation strategies such as bitwise/bytewise flips, block replacement, and crossover on two seed files; (3) running the target program against the newly generated inputs, and recording the executed traces; (4) reporting fault seeds if crashes are detected, and saving those interesting seeds that cover new traces into the seed pool. Such iteration continues until given computation resource exhausts. The two key components in CGF are {\it mutation} and {\it coverage feedback} that largely determine the efficiency of fuzzing.

Despite the huge differences between traditional programs and DNNs, the success of CGF on the former still gains insight into the fuzzing on the latter. For example, the target traditional program mirrors the DNN, the seed of fuzzer mirrors the input of the DNN, and the coverage feedback could be some coverage of DNN. Considering the unique characteristics of the DNN, it is still challenging to develop effective {\it mutation strategies} and {\it coverage criteria} in terms of DNN fuzzing. This paper aims to fill this gap by designing effective CGF framework towards providing a quality assurance gadget during the DNN development and deployment process.

\section{METHODOLOGY}
In this section, we elaborate proposed coverage guided fuzzing for DNNs. We take an overview of {\tool} and then describe each of the key components in details.
\subsection{Overview of DeepHunter}
\begin{figure*}
    \centering
	\includegraphics[width=\textwidth]{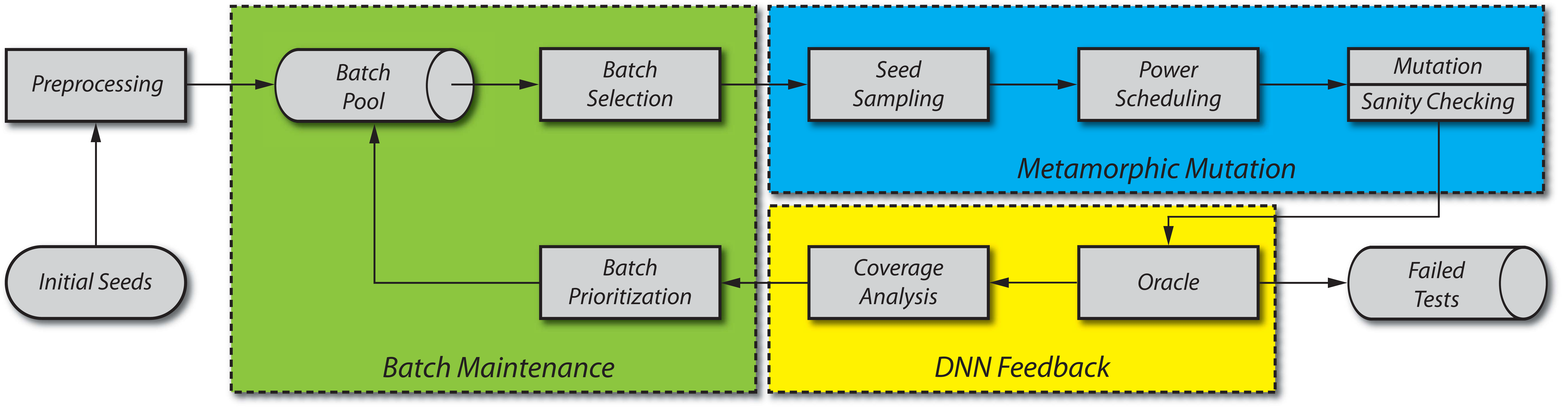}
	\caption{The workflow of {\tool}, which leverages metamorphic mutation to generate tests with coverage feedback as guidance.}
	\label{fig:overview}
	\vspace{-3mm}
\end{figure*}

Fig.~\ref{fig:overview} depicts the overview of {\tool}, and Algorithm~\ref{algo1} specifies the details. At a high level, {\tool} consists of three major components: {\it Metamorphic Mutation}, {\it DNN Feedback}, and {\it Batch Pool Maintenance}. We define an atomic input of the DNN (\eg, an image) as a \emph{seed} and a set of seeds (\eg, multiple images) as a \emph{batch}. As DNNs can quickly predict multiple seeds (\ie, a batch) at once, we maintain a batch pool instead of a seed pool to improve the fuzzing effectiveness. During fuzzing, {\tool}~first selects a batch and generates a large number of mutated seeds, then the DNN predicts all mutated seeds \emph{at once}. At last, \tool~maintains the pool based on the coverage information. The workflow of {\tool} is detailed below (see Algorithm~\ref{algo1}).

The inputs of \tool~are the initial seeds and the target DNN model under test. Before the fuzzing loop, initial seeds are constructed as batches, which are added into the batch pool (Line~\ref{algo1:preprocess}). During fuzzing process, the fuzzer selects one batch from the priority batch pool (Line~\ref{algo1:select}). From the batch, the fuzzer samples some seeds to be mutated (Line~\ref{algo1:sample}). The fuzzer applies a power scheduling against the sampled seeds to determine the mutation chances for each seed (\cf Section~\ref{sec3:powerschedule}). 
For each sampled seed, the fuzzer will mutate it for the assigned times (Line~\ref{algo1:mutationnumber}), and sanitize each mutated seed (\cf  Section~\ref{sec3:mutation}) since random mutation may generate some meaningless seeds (\eg, images imperceptible to the human eye). For each valid mutant of the original seeds, the test oracle will verify whether this is a failed test. After mutating all sampled seeds, the survived mutants are constructed as a batch. DNNs will predict all the seeds and collect the coverage information of the batch (Line~\ref{algo1:predictall}). If the batch gains the coverage, it will be added into the batch pool. Batch prioritization will prioritize the batches that have been seldomly fuzzed (Lines~\ref{algo1:addtopool}-\ref{algo1:prioritize}, \cf Section~\ref{sec3:coverage} and Section~\ref{sec3:seedpriorization}).

\begin{algorithm}[t]
% 	\small
	\footnotesize
	\SetKwInOut{Input}{input}\SetKwInOut{Output}{output}
	\SetKwInOut{Const}{const}
	\Input{$I$: Initial Seeds, $DNN$, Target Neural Network}
	\Output{$F$: Failed Tests}
	\Const{$K$: Total number of mutation for a batch}
% 	\BlankLine
	$F \leftarrow \emptyset$\;
	$T \leftarrow Preprocess(I)$\; \label{algo1:preprocess}
	\While{$B \leftarrow SelectNext(T)$}{ \label{algo1:select}
		$S \leftarrow  Sample(B)$\;\label{algo1:sample}
		$P_S \leftarrow  PowerSchedule(S, K)$\; \label{algo1:powerschedule}
		$B' \leftarrow \emptyset$\;
		\For{$\image \in S$}{
			\For{$i$ from $1$ to $P_S(\image)$}{\label{algo1:mutationnumber}
				$\image' \leftarrow Mutate(\image, B)$\;
				\If{isFailedTest($\image'$)}{
					$F \leftarrow F \bigcup \{\image'\}$\;
				}\ElseIf{isChanged($\image,\image'$)}{
					$B' \leftarrow B' \bigcup \{\image'\}$\;
				}
			}
		}
		$cov \leftarrow Predict(DNN, B')$\;\label{algo1:predictall}
		\If{$CoverageGain(cov)$}{
			$T \leftarrow T \bigcup \{B'\}$\;\label{algo1:addtopool}
			$BatchPrioritize(T)$\;\label{algo1:prioritize}
		}
	}
	\caption{\tool}\label{algo1}
\end{algorithm}

\subsection{Transformation and Mutation}\label{sec3:mutation}
Traditional fuzzers such as AFL mutate the input with bitwise / bytewise flips, block replacement, crossover between input files, \etc. However, these strategies usually generate too many inputs that are meaningless in DNN fuzzing. For example, images or voices that are imperceptible to human senses should be discarded from the mutation. Hence, one challenge is how to balance between increasing the changeability of mutation and generating meaningful inputs. If the mutation change is very small, the newly generated input may be almost unchanged; despite the fact that it may be meaningful, the fuzzer has lower chances of finding failed tests. On the other hand, if the mutation change is very large, more failed tests may be identified; however, the failed tests are more likely to be meaningless.

In this work, we mainly focus on image inputs.
To solve the aforementioned challenge, we develop a metamorphic mutation strategy. The basic objective is that {\it given an image~$i$, the mutator generates another new image $i'$ such that the semantics of $i$ and $i'$ are the same from the perspective of people}.

\noindent\textbf{Image Transformation.} To increase  the changeability of mutation, we select eight image transformations which are classified into two categories:
\begin{itemize}[wide=1pt,leftmargin=10pt]
	\item {\it Pixel Value transformation} $\pixeltransform$: change image contrast, image brightness, image blur and image noise.
	\item {\it Affine transformation} $\affinetransform$: image translation, image scaling, image shearing and image rotation.
\end{itemize}

Intuitively, {\it Pixel Value transformation} changes the pixel values of the image while {\it Affine transformation} moves the pixels of the image. The transformations have been proved to be effective and useful in~\cite{tian2018deeptest}.

\begin{definition}\label{def:mutation}
	An image $\image'$ is {\it one-time mutated} from $\image$ if $\image'$ is generated after a transformation $t$ on $\image$ (denoted as $\image \xlongrightarrow {t} \image'$), where $t \in \pixeltransform \bigcup \affinetransform$. An image $\image'$ is {\it sequentially mutated} from $\image$ if $\image'$ is generated after a sequence of one-time mutations ($\image\xlongrightarrow {t_0} \image_1, \image_1\xlongrightarrow {t_1} \image_2, \ldots, \image_n \xlongrightarrow {t_n} \image'$) (denoted as $\image\xlongrightarrow {t_0,t_1, \ldots,t_n}\image'$). 
\end{definition}

\noindent \textbf{Metamorphic Mutation.} By setting proper parameters for different transformations, it is assumed that the image after {\it one-time} mutation has the same semantics with the original image. However, during fuzzing, one image can be sequentially mutated from the original image, it is challenging to generate meaningful images after a sequence of mutations. To boost the mutation effectiveness, we propose the metamorphic mutation.

In order to ensure the meaningfulness of the mutated image as much as possible, we adopt a conservative strategy that makes {\it Affine Transformation} to be selected only once because multiple affine transformations are more likely to generate meaningless images. We assume that an affine transformation will not affect the semantics under the selected parameters. {\it Pixel Value Transformation} can be selected multiple times and we use $L_0$ and $L_\infty$ to limit the pixel-level change. Suppose an image $\image$ is mutated to $\image'$ by a pixel value transformation, then $\image'$ is meaningful in terms of $\image$ if $f(\image,\image')$ (Equation~\ref{equa:mutation}) is satisfied.
\begin{equation}
\label{equa:mutation}
f(\image,\image') =
\left\{
\begin{array}{lr}
L_\infty \le 255, & \mathrm{if}~L_0 < \alpha \times \mathrm{size}(\image) \\
L_\infty < \beta \times 255, &  \mathrm{otherwise} 
\end{array}
\right.
\end{equation}
where $0<\alpha, \beta< 1$, $\mathit{L_0}$ represents the maximum number of the changed pixels, $\mathit{L_\infty}$ represents the maximum value of the pixel changes, $\mathrm{size}(\image)$ is the number of pixels in image $0< \image$.

Intuitively, if the number of changed pixels is very small ($< \alpha \times \mathrm{size}(\image)$), we assume it does not change the semantics and $L_\infty$ can be any value. If the number of changed pixels exceeds the boundary, we limit the maximum changed value ($< \beta \times 255$).

\begin{algorithm}[t]
% 	\small
    \footnotesize
	\SetKwInOut{Input}{input}\SetKwInOut{Output}{output}
	\SetKwInOut{Const}{const}
	\Input{$\image$: Seed}
	\Output{$\image_0'$: New Seed}
	\Const{$TRY\_NUM$: The maximum number of trials}
% 	\BlankLine

	$(\image_0, \image_0', state) \leftarrow info(\image)$\; \label{algo2:gettuple}
	\For{$i$ from 1 to $TRY\_NUM$}{\label{algo2:beginmutate}
		\If{$state == 0$}{
			$t \leftarrow randomPick(\affinetransform\bigcup\pixeltransform)$\;\label{algo2:pickpa}
		}\Else{
			$t \leftarrow randomPick(\pixeltransform)$\;\label{algo2:pickp}
		}
		$p \leftarrow pickRandomParam(t)$\; \label{algo2:pickparameter}
		$\image' \leftarrow t(\image, p)$\;\label{algo2:mutation}
		\If{$isSatisfied\big(f(\image_0',\image')\big)$}{\label{algo2:checkmean}
			\If{$t \in \affinetransform$}{
				$state \leftarrow 1$\;\label{algo2:setone}
				$\image_0' \leftarrow t(\image_0, p)$\;\label{algo2:updateori}
			}
			$info(\image') \leftarrow (\image_0, \image_0', state)$\;\label{algo2:save}
			\Return $\image'$\;\label{algo2:ends}
		}
	}\label{algo2:endmutate}
	\Return $\image$\;
	\caption{Mutate}\label{algo2}
\end{algorithm}

\begin{definition}\label{def:ori}
	Given a mutated image $\image$, the {\it original image} (denoted as $\image_0$) of $\image$ is the image in the initial seeds and $\image$ is one-time mutated or sequence mutated from $\image_0$, \ie, $\image_0 \xlongrightarrow{t_0,\ldots,t_n}\image$, where $n\ge 0$. The {\it reference image} (denoted as $\image_0'$) is defined as:
	% ----------------------------
	\begin{equation*}
	\label{equa:reference}
	\image_0' =
	\left\{
	\begin{array}{lr}
	\image_j, & \exists~0\le j\le n.~t_j\in\affinetransform \wedge \image_0\xlongrightarrow{t_0,\ldots,t_j}\image_j \\
	\image_0, &  \mathrm{otherwise} 
	\end{array}
	\right.
	\end{equation*}
	% ----------------------------
\end{definition}

Algorithm~\ref{algo2} shows the details of the mutation, which takes an original image $\image$ as the input and the mutated image $\image'$ as the output. We first obtain the tuple $(\image_0, \image_0',state)$ (Line~\ref{algo2:gettuple}) which is recorded in the batch. $state$ is the current mutation state $0$ or $1$, which represents whether an {\it Affine Transformation} is used. {\tool} tries to mutate a meaningful image $\image'$ with a maximum number of trials $TRY\_NUM$ (Line~\ref{algo2:beginmutate}-\ref{algo2:endmutate}). It randomly picks a transformation $t$. If the current mutation state is $0$, it can select from both  {\it Affine Transformation} and  {\it Pixel Value Transformation} (Line~\ref{algo2:pickpa}). If the mutation state is $1$, it can only use a pixel value transformation (Line~\ref{algo2:pickp}). For the transformation $t$, it picks a parameter randomly (Line~\ref{algo2:pickparameter}) and performs the transformation (Line~\ref{algo2:mutation}). 

Next, Algorithm~\ref{algo2} computes $L_0$ and $L_\infty$ between the reference image $\image_0'$ and the new mutated image $\image'$ to check whether $\image'$ is meaningful (Line~\ref{algo2:checkmean}). Note that we compare $\image'$ with reference image $\image_0'$ instead of original image $\image_0$ because: (1) the pixels between $\image_0'$ and $\image'$ are corresponding, which is necessary to compute $L_0$ and $L_\infty$ and (2) we assume that $\image_0'$ and $\image_0$ have the same semantics under our conservative parameters. Hence, if $f(\image_0',\image')$ (\cf Equation~\ref{equa:mutation}) is satisfied, we can conclude that $\image'$ and $\image_0$ also have the same semantics and the mutation is successful. If the selected $t$ is an {\it Affine Transformation}, it updates the mutation state of $\image'$ and $\image_0'$ (Line~\ref{algo2:setone}-\ref{algo2:updateori}). At last, Algorithm~\ref{algo2} saves the current image $\image'$ (Line~\ref{algo2:save}) and ends the mutation (Line~\ref{algo2:ends}). If there is no successful mutation after $TRY\_NUM$ mutations, it outputs the image $\image$. 

\subsection{Power Scheduling}\label{sec3:powerschedule}

As described in Algorithm~\ref{algo2}, {\tool} mutates one image with a limited number of tries. If $f(\image_0', \image')$ is satisfied, the mutation of $\image$ is successful. Actually, the possibility of successful mutation depends on the difficulty that $f(\image_0', \image')$ is satisfied.

	Given an image $\image$ and its reference image $\image_0'$, we define its {\it mutation potential} as $\beta \times 255 \times \mathrm{size}(\image) - \mathrm{sum}(\mathrm{abs}(\image-\image_0')).$ 
Intuitively, the mutation potential approximately represents the mutation space of an image $I$, \ie, the difficulty that $f(\image_0', \image')$ is satisfied. $\beta \times 255 \times \mathrm{size}(\image)$ represents the maximum value that the image can change. $\mathrm{sum}(\mathrm{abs}(\image-\image_0'))$ represents the value that the image has changed in terms of $\image_0'$. For example, suppose $\image'$ is sequentially mutated from $\image_0'$ : $(\image_0'\xlongrightarrow{t_0}\image_1,\ldots, \image_n \xlongrightarrow{t_n} \image')$, the mutation potential of images in the front of the sequence (\eg, $\image_1$) is more likely to be higher than those  in the tail (\eg, $\image'$).

The power scheduling is a procedure for {\tool} to decide mutation chances for different seeds (\ie, images). To boost the efficiency of fuzzing, we expect to mutate more images that have higher mutation potential.

%%%%%%%%%%%%%%%%%%%%%%%%%%%%%%%%%%%%%%%%%%%%%%%%%%%%%%%%%%%%%%%%%%%%%%%%%%%%%%%%%%%%%%%%%%%%%%%

\subsection{Plugable Coverage-Guided Fuzzing} 
\label{sec3:coverage}

A dumb fuzzer without any coverage guidance aimlessly mutates the seed, without knowing whether the generated test input is preferable. Consequently, such a fuzzer may frequently keep seeds that do not bring new desired information; even worse, mutation on these seeds may bury other ``interesting'' seeds in the fuzzing queue, significantly decreasing the fuzzing effectiveness. Therefore, modern fuzzers for traditional software often embrace some feedback such as \emph{code coverage}. 

In this paper, {\tool} selects six different criteria~\cite{pei2017deepxplore,ma2018deepgauge} (Table~\ref{tab:model_summary}) as different feedback to determine whether the newly generated batch should be kept for further mutation. The criteria have been proven to be useful to capture the internal DNN states. However, due to the huge numerical space of each neuron value and the large scale nature of the DNN software, the fuzzer could be overloaded with \emph{flooded feedback}.  In fact, without triaging, seed inputs with similar neuron values will be \emph{unnecessarily} retained. Due to the mutation instinct, there will be a huge number of such mutants that originate from a given seed. To tackle this issue, we equally split the numerical neuron-feedback interval of each criteria into different buckets, each of which will be regarded as an ``equivalent class''. If a new seed with its coverage results of a neuron falling into existing buckets, it is out of the interest and discarded. This mechanism is inspired from the ``loop bucket'' practice used in the traditional fuzzing framework~(\eg, AFL), to mitigate the trace exploitation issue~\cite{afl_detail}.

\begin{table}[!t]
 \vspace{2mm}
\scriptsize
\caption{The plugable coverage criteria integrated in {\tool} for test guidance. Besides the first five criteria originally proposed in~\cite{pei2017deepxplore,deepguage}, we also include an extra BKNC criterion that plays as a counterpart of TKNC by measuring the ratio of top-k most hypoactived neurons.}
%\vspace{-2mm}
\begin{center}
\begin{tabular}{ll}
\hline 
\textbf{Subject Cov. Criteria} & \textbf{Description}\tabularnewline
\hline 
\textbf{Neuron Cov. (NC)} & The ratio of activated neurons\tabularnewline
\textbf{K-multisec. Neu. Cov. (KMNC)} & The ratio of covered k-multisections of neurons\tabularnewline
\textbf{Neuron Bound. Cov. (NBC)} & The ratio of covered boundary region of neurons\tabularnewline
\textbf{Strong Neuron Act. Cov. (SNAC)} & The ratio of covered hyperactive boundary region\tabularnewline
\textbf{Top-k Neu. Cov. (TKNC)} & The ratio of neurons in top-k hyperactived state\tabularnewline
\textbf{Bottom-k Neu. Cov. (BKNC)} & The ratio of neurons in top-k hypoactived\tabularnewline
\hline 
\end{tabular}

\end{center}
\label{tab:model_summary}
\vspace{-2mm}
\end{table}

\subsection{Batch Prioritization}
\label{sec3:seedpriorization}
Batch prioritization decides which batch should be picked next. We adopt a strategy which probabilistically selects the batch based on the number of times it has been fuzzed. Specially, the probability is computed by:
%-------------------------------
\begin{equation}
\label{equa:probmutation}
P(B) =
\left\{
\begin{array}{lr}
1-f(B)/\gamma,  & \mathrm{if}~f(B) < (1-p_{\min}) \times \gamma  \\
p_{\min} , &  \mathrm{otherwise} 
\end{array}
\right.
\end{equation}
%-------------------------------
where $B$ is a batch, $f(B)$ represents how many times the batch $B$ has been fuzzed and $p_{\min} > 0$ is the minimum probability. The values of parameters $\gamma$ and $p_{\min} $ can be adjusted. 

The basic idea here is to prioritize the batches that have been seldomly fuzzed. For example, the probability of new mutated batch is 1 since it gains new coverage and is regarded as interesting. To keep the diversity, other batches that have been fuzzed many times also have a minimum probability $p_{\min}$ to be selected.

\label{sec:mutation_testing}

\section{Experiments}
\label{sec:exp}

{\tool} is implemented in Python and C: the metamorphic mutation component and DNN coverage feedback component are implemented in Python based on deep learning framework Keras ~(ver.2.1.3)~\cite{keras} with TensorFlow~(ver.1.5.0) backend~\cite{tensorflow}; the batch maintenance component is implemented in C for efficiency. 
We evaluate {\tool} by investigating the following research questions:

\begin{itemize}[leftmargin=9mm]

\item [\textbf{RQ 1:}]
What coverage can {\tool} achieve when guided by the six testing criteria?

\item [\textbf{RQ 2:}]
Does {\tool} facilitate the DNN model evaluation effectively?

\item [\textbf{RQ 3:}] Can {\tool} enable diverse erroneous behavior detection of DNNs?

\item [\textbf{RQ 4:}] Can {\tool} detect potential defects introduced during DNN quantization?

\end{itemize}

\begin{table}[!t]
\vspace{2.2mm}

% \scriptsize

\setlength\tabcolsep{3.0pt}
\caption{Subject datasets and DNN models.}
\vspace{-2mm}
\begin{center}
\begin{tabular}{lccccl}
\hline 
\multirow{2}{*}{\textbf{DataSet}} & \textbf{Dataset} & \multirow{2}{*}{\textbf{DNN Model}} & \multirow{2}{*}{\textbf{\#Neuron}} & \multirow{2}{*}{\textbf{\#Layer}} & \textbf{Test}\tabularnewline
 & \textbf{Description} &  &  &  & \textbf{Acc.}\tabularnewline
\hline 
\multirow{3}{*}{\textbf{MNIST}} & Hand written & LeNet-1 & 52 & 7 & 0.976\tabularnewline
 & digits recog. & LeNet-4 & 148 & 8 & 0.989\tabularnewline
 & from 0 to 9 & LeNet-5 & 268 & 9 & 0.990\tabularnewline
\hline 
\multirow{2}{*}{\textbf{CIFAR-10}} & General image & ResNet-20 & 2,570 & 70 & 0.917\tabularnewline
 & with 10-class & VGG-16 & 12,426 & 17 & 0.928\tabularnewline
\hline 
\multirow{2}{*}{\textbf{ImageNet}} & 1000-class large  & MobileNet & 38,904 & 87 & 0.871$^*$\tabularnewline
 & scale image cla. & ResNet-50 & 94,059 & 176 & 0.929$^*$\tabularnewline
\hline 
\end{tabular}

\end{center}
\vspace{-1mm}

\label{tab:model_summary_1}
\vspace{-2mm}
\begin{center}
\scriptsize{* The reported top-5 test accuracy of pretrained DNN model in~\cite{keras-model}.}
\end{center}
\vspace{-2mm}
\end{table}

%-----------------------------------

\begin{figure}[t]
\centering
\includegraphics[width=0.95\linewidth]{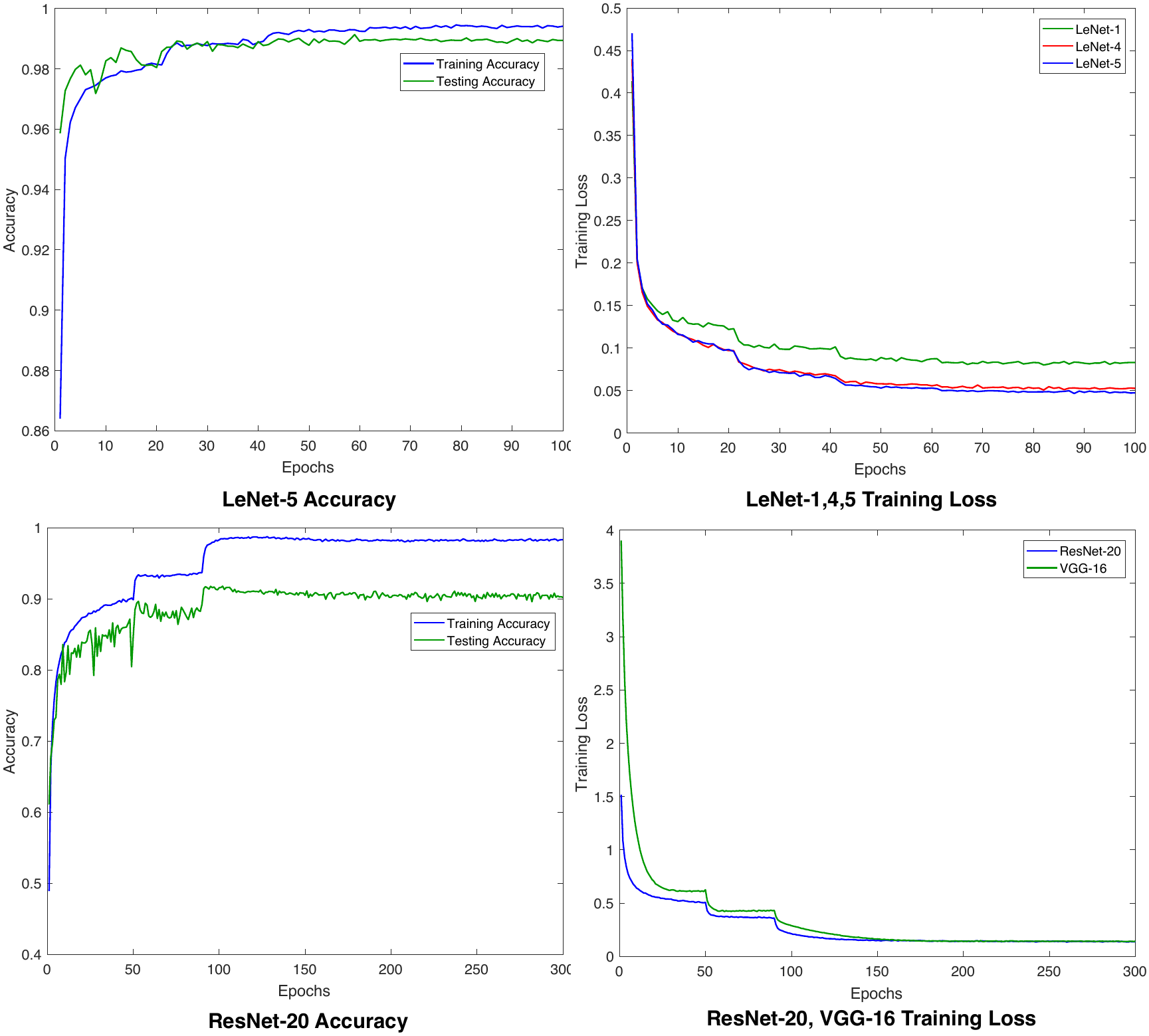}

\caption{Controlled results on training loss, training accuracy, and test accuracy. 
}
\vspace{-2mm}
\label{fig:train_plot}

\end{figure}

\subsection{Datasets and DNN Models}

We select three popular publicly available datasets (\ie, MNIST~\cite{mnist}, CIFAR-10~\cite{cifar}, and ImageNet~\cite{imagenet}) as the evaluation subject datasets~(see Table~\ref{tab:model_summary_1}). For each dataset, we study popular DNN models~\cite{lenet-fam,2018arXiv180102610X,carlini2017towards} that are widely used in previous work.
In particular, we perform extensive controlled study on MNIST and CIFAR-10, and investigate the scalability and usefulness of {\tool}. 
Table~\ref{tab:model_summary_1} summarizes the structures and complexity of the DNNs used in this paper.

\noindent\textbf{MNIST} is for handwritten digit image recognition, containing $60,000$ training data and $10,000$ test data, with a total number of $70,000$ data in 10 classes~(\ie, handwritten digits from $0$ to $9$). Each MNIST image is a single-channel of size $28 \times 28 \times 1$.
On MNIST, we have studied three LeNet family models~(LeNet-1, LeNet4, LeNet-5~\cite{lenet-fam}) as the subject models. We train each of the LeNet models in a controlled setting, with detailed configuration and discussion in Section~\ref{subsec:rq2}.

\noindent\textbf{CIFAR-10} is a collection of images for general-purpose image classification, including $50,000$ training data and $10,000$ test data in $10$ different classes~(\eg, airplanes, cars, birds, and cats). Each CIFAR-10 image is three-channel of size $32 \times 32 \times 3$. 
The classification task of CIFAR-10 is generally harder  than that of MNIST due to the data size and complexity.
To obtain competitive performance on CIFAR-10, we study two well-known DNN models~(\ie, ResNet-20~\cite{resnet} and VGG16~\cite{vgg}) as the subject models.

\noindent\textbf{ImageNet.} To further demonstrate that {\tool} scales to practical-sized dataset and DNN models, we also select ImageNet, which is a large-scale visual recognition challenge~(ILSVRC) dataset for general-purpose image classification. The complexity of ImageNet is characterized by a large number of training data~(\ie, over one million) and test data~(\ie, 50,000), as well as large data points, each of which is of size $224 \times 224 \times 3$ ($\sim 50$x dimensionality of CIFAR-10). Therefore, it would be an ordeal for any automated testing tool to work on ImageNet sized dataset and DNN models. Specifically, we try to examine whether {\tool} enables the fuzz testing on ImageNet dataset and practical-sized DNN models~(\ie, VGG-19~\cite{vgg}, ResNet-50~\cite{resnet}).

\begin{table}[!t]
\vspace{2mm}
\centering
\scriptsize

\setlength\tabcolsep{3.0pt}
\caption{DNN model and their training and test performance.}
\vspace{-2mm}
\begin{center}
\begin{tabular}{ccccccc}
\hline 
\textbf{DataSet} & \textbf{DNN} & \textbf{Epoch} & \textbf{Syno.} & \textbf{Train Loss} & \textbf{Train Acc.} & \textbf{Test Acc.}\tabularnewline
\hline 
 &  & 10 & A & 0.131 & 0.965 & 0.967\tabularnewline
 & \textbf{LeNet-1} & 30 & B & 0.099 & 0.975 & 0.975\tabularnewline
 &  & 45 & C & 0.087 & 0.979 & 0.976\tabularnewline
\cline{2-7} 
 &  & 10 & A & 0.117 & 0.974 & 0.978\tabularnewline
\textbf{MNIST} & \textbf{LeNet-4} & 25 & B & 0.077 & 0.986 & 0.986\tabularnewline
 &  & 50 & C & 0.058 & 0.990 & 0.989\tabularnewline
\cline{2-7} 
 &  & 10 & A & 0.116 & 0.977 & 0.983\tabularnewline
 & \textbf{LeNet-5} & 30 & B & 0.071 & 0.988 & 0.989\tabularnewline
 &  & 45 & C & 0.056 & 0.992 & 0.990\tabularnewline
\hline 
 &  & 40 & A & 0.515 & 0.894 & 0.859\tabularnewline
 & \textbf{ResNet-20} & 55 & B & 0.385 & 0.932 & 0.880\tabularnewline
\multirow{2}{*}{\textbf{CIFAR-10}} &  & 95 & C & 0.239 & 0.977 & 0.917\tabularnewline
\cline{2-7} 
 &  & 30 & A & 0.623 & 0.914 & 0.850\tabularnewline
 & \textbf{VGG-16} & 55 & B & 0.443 & 0.965 & 0.900\tabularnewline
 &  & 95 & C & 0.316 & 0.995 & 0.928\tabularnewline
\hline 
\end{tabular}
\end{center}
\label{tab:controlled_model_summary}
\vspace{-4mm}
\end{table}

%-----------------------------------

\subsection{Experiment Setup}

\noindent\textbf{Subject DNN Model Training and Preparation.}
Since the DNN model quality could affect the evaluation results, we carefully select the well-known DNN models that obtain competitive performance on each studied dataset. In this paper, we closely follow the common machine learning training practice and instructions~\cite{krizhevsky2012imagenet,goodfellow2016deep}, and set up a three-stage adaptive learning rate for training MNIST and CIFAR-10 DNN models. 
The larger learning rate at the early training stage accelerates the training convergence, while the later stage with a smaller learning rate allows the performance fine-tuning. The training loss, training accuracy, and test accuracy for each model are shown in Figure~\ref{fig:train_plot}. From the training accuracy and training loss curve, we can observe that the training process jumps into three different stages as expected. We follow the machine learning practice and select the best candidate models with the most competitive performance without overfitting as the subject DNN model instances for fuzz testing (see~Table~\ref{tab:model_summary} and DNN C variants in Table~\ref{tab:controlled_model_summary} with the detailed epochs and training information).
Due to the large size of training data and training effort of ImageNet, we select the pretrained MobileNet and ResNet-50~\cite{keras-model} as the subject models~\cite{resnet}.

For RQ 2, we try to evaluate whether {\tool} enables the DNN model quality evaluation. Therefore, besides the best candidate C instances for each model used for MNIST and CIFAR-10, we also select other two instances A and B from each of the first two training stages, which allows us to sort the model quality relation $Q_A < Q_B < Q_C$ as the groundtruth.\footnote{\scriptsize The general DNN quality groundtruth is hard to obtain; the state of the practice still relies on test accuracy. 
Our three-stage training procedure follows the machine learning practice~\cite{krizhevsky2012imagenet,goodfellow2016deep} and allows to obtain model instances, each from one of the three training stages, so that we could obtain the desired model quality relation with high confidence.
}
The selected groundtruth model instances for quality evaluation is summarized in Table~\ref{tab:controlled_model_summary}.

\noindent\textbf{Coverage-Guided Fuzz Testing.}
After all DNN models are obtained for each dataset~(see Table~\ref{tab:model_summary_1}), we use {\tool} to perform large-scale fuzz testing on each of these models to generate tests. For each DNN model,
we randomly sample tests as initial seed batches from their original test dataset such that all these tests are correctly handled by the model.\footnote{\scriptsize We sample 1,000 initial seeding data in 10 batches (each contains 100 test data) for MNIST and CIFAR-10, and 500 seeding data for ImageNet in 20 batches w/ equal size.} Furthermore, on MNIST and CIFAR-10 datasets, the sampled initial seeds are also correctly handled by each of the model instance A and B, as shown in Table~\ref{tab:controlled_model_summary}. This allows us to perform the controlled evaluation~(for RQ 2) on the usefulness of {\tool} for quality evaluation of models, where the initial seeds fail to distinguish each of the model instances A, B, C in terms of prediction accuracy.
We configure {\tool} to use the each of the six studied testing criteria for guided fuzz testing with a time budget of $24$ hours, allowing the testing coverage to achieve relatively saturated status. The obtained tests are used to perform post-phase analysis for each research question. 
Note that the post-phase analysis for all the research questions are also computational intensive.
To support such a large-scale subject DNN model set training, which are fuzz testing and post-phase analysis, we run all the experiments on a high performance computer cluster. Each cluster node runs a GNU/Linux system with Linux kernel 4.4.0 on a 28-core 2.3GHz Intel Xeon 64-bit CPU with 196 GB of RAM equipped with a NVIDIA Tesla V100 16G-GPU.

\subsection{Coverage Results Guided by Different Testing Criteria}
\label{subsec:rq1}
\textbf{To answer RQ 1}, the achieved coverage of {\tool} guided by different testing criteria is shown in Table~\ref{tab:coverage_result}. {\tool} generates tests that significantly increase the corresponding coverage compared with initial seeds, as confirmed by Wlicoxon Singed Ranks Test ($p<0.01$) for all cases. 

The difficulty to cover different criteria is different. For example, the TKNC obtained by initial seeds has already obtained very high coverage in some cases~(see LN-5 with $76.3\%$ initial NC), While the initial seeds only obtain $0.1\%$ for ResNet-20 on NBC and SNAC. Such results are consistent with the coverage trends reported in~\cite{deepguage}, where the NBC and SNAC are more challenging to cover since they represent the corner-regions where neuron states go beyond a normal region. Even though, {\tool} is still able to significantly boost such criteria, increased by 15.86x~(from $0.7\%$ to $10.3\%$ on MobileNet) to 77.5x (from $0.04\%$ to $3.1\%$ on VGG-16).
\vspace{2mm}
\begin{tcolorbox}[size=title]
{ \textbf{Answer to RQ 1:} {\tool} significantly boosts the coverage across with different criteria guidance.} 
\end{tcolorbox}

\begin{table}[!t]
\vspace{2mm}
\centering
\scriptsize
\setlength\tabcolsep{2.6pt}
\caption{The coverage of initial seeds and  tests with {\tool} guided by the corresponding testing criteria.}
\vspace{-2mm}
\begin{center}
\begin{tabular}{lcccccccccccc}
\hline 
\textbf{DNN} & \multicolumn{2}{c}{\textbf{NC(\%)}} & \multicolumn{2}{c}{\textbf{KMNC(\%)}} & \multicolumn{2}{c}{\textbf{NBC(\%)}} & \multicolumn{2}{c}{\textbf{SNAC(\%)}} & \multicolumn{2}{c}{\textbf{TKNC(\%)}} & \multicolumn{2}{c}{\textbf{BKNC(\%)}}\tabularnewline
\textbf{Model} & \textbf{Init.} & \textbf{D.H.} & \textbf{Init.} & \textbf{D.H.} & \textbf{Init.} & \textbf{D.H.} & \textbf{Init.} & \textbf{D.H.} & \textbf{Init.} & \textbf{D.H.} & \textbf{Init.} & \textbf{35.4}\tabularnewline
\hline 
\textbf{LN-1} & 22.9 & 31.3 & 31.4 & 92.2 & 1.2 & 24.4 & 0.3 & 22.1 & 49.7 & 49.8 & 49.8 & 49.8\tabularnewline
\textbf{LN-4} & 56.3 & 61.1 & 21.9 & 75.9 & 0.4 & 15.1 & 0.2 & 20.2 & 69.5 & 72.7 & 27.2 & 31.1\tabularnewline
\textbf{LN-5} & 58.0 & 70.8 & 20.6 & 78.4 & 0.3 & 11.3 & 0.2 & 19.2 & 76.3 & 83.2 & 24.6 & 30.5\tabularnewline
\textbf{RN-20} & 7.5 & 10.8 & 36.0 & 75.1 & 0.1 & 8.1 & 0.1 & 8.11 & 62.3 & 68.0 & 63.3 & 68.7\tabularnewline
\textbf{VGG16} & 41.9 & 46.2 & 41.1 & 84.1 & 0.04 & 3.1 & 0.1 & 3.1 & 13.3 & 15.1 & 13.9 & 16.2\tabularnewline
\textbf{MN} & 7.0 & 7.8 & 26.9 & 73.4 & 0.7 & 10.3 & 0.4 & 9.5 & 4.9 & 7.9 & 5.2 & 8.5\tabularnewline
\textbf{RN-50} & 4.8 & 9.4 & 23.9 & 51.6 & 0.1 & 3.2 & 0.1 & 3.8 & 14.1 & 22.9 & 13.7 & 20.6\tabularnewline
\hline 
\textbf{Avg.} & 28.3 & 33.9 & 28.8 & 75.8 & 0.4 & 10.8 & 0.2 & 12.3 & 41.4 & 45.6 & 28.2 & 32.2\tabularnewline
\hline 
\end{tabular}

\end{center}
\label{tab:coverage_result}
 \vspace{-2mm}
\end{table}

%-----------------------------------

\subsection{DNN Model Quality Evaluation}
\label{subsec:rq2}
 Although software quality standards and metrics are well-established for traditional software, the DNN quality assurance research is still at an early stage, with most current work relying on test accuracy. \textbf{To answer RQ 2}, we investigate whether {\tool} provides more useful feedback and evaluation on the DNN model quality, by using controlled setting on MNIST and CIFAR-10~(see Table~\ref{tab:controlled_model_summary}). Note that the initial seeds are all correctly predicted by all instances A, B, C of each model.

We have used {\tool} to generate tests for instance C of each model. Among these tests, we keep those correctly predicted by C, and run these tests on instances A and B  of each model. The obtained accuracy for instances A and B of each model is shown in Table~\ref{tab:model_quality}. 
We can see that {\tool} facilitates the DNN model quality evaluation, and the quality evaluation results are largely consistent with the model quality groundtruth~(\ie, $Q_B > Q_A$). The tests generated by different coverage guidance exhibit different abilities to show the model quality difference. For example, on LeNet-5, the NC only slightly shows B might have better quality; however, this becomes obvious when it comes to NBC, where instance A achieves $88.4\%$ accuracy and instance B achieves $96.9\%$.

\begin{table}[!t]
%\vspace{2mm}
\centering
\scriptsize
\caption{The controlled DNN model instance quality evaluation accuracy results.}
\vspace{-2mm}
\begin{center}
\begin{tabular}{lcccccccccccc}
\hline 
\textbf{DNN} & \multicolumn{2}{c}{\textbf{NC(\%)}} & \multicolumn{2}{c}{\textbf{KMNC(\%)}} & \multicolumn{2}{c}{\textbf{NBC(\%)}} & \multicolumn{2}{c}{\textbf{SNAC(\%)}} & \multicolumn{2}{c}{\textbf{TKNC(\%)}} & \multicolumn{2}{c}{\textbf{BKNC(\%)}}\tabularnewline
\textbf{Instan.} & \textbf{A} & \textbf{B} & \textbf{A} & \textbf{B} & \textbf{A} & \textbf{B} & \textbf{A} & \textbf{B} & \textbf{A} & \textbf{B} & \textbf{A} & \textbf{B}\tabularnewline
\hline 
\textbf{LN-1} & 99.0 & 99.5 & 92.8 & 97.5 & 93.0 & 96.5 & 90.1 & 96.0 & 91.5 & 98.5 & 95.4 & 98.7\tabularnewline
\textbf{LN-4} & 98.1 & 99.6 & 92.7 & 97.3 & 87.1 & 95.5 & 91.0 & 95.4 & 91.5 & 95.7 & 89.3 & 95.9\tabularnewline
\textbf{LN-5} & 95.4 & 97.7 & 91.3 & 94.1 & 88.4 & 96.9 & 91.1 & 97.0 & 92.4 & 97.0 & 91.3 & 96.7\tabularnewline
\textbf{RN-20} & 92.6 & 93.9 & 81.7 & 87.2 & 83.8 & 87.4 & 83.8 & 86.3 & 81.9 & 87.1 & 85.3 & 87.8\tabularnewline
\textbf{VGG16} & 86.0 & 87.7 & 78.0 & 83.3 & 80.0 & 81.3 & 82.9 & 83.9 & 82.8 & 83.3 & 84.1 & 85.2\tabularnewline
\hline 
\textbf{Avg.} & 94.2 & 95.7 & 87.3 & 91.9 & 86.4 & 91.5 & 87.8 & 91.7 & 88.0 & 92.3 & 89.1 & 92.9\tabularnewline
\hline 
\end{tabular}

\end{center}
%\vspace{-2mm}
\label{tab:model_quality}
 \vspace{-4mm}
\end{table}

We see that most of the accuracy of instances A and B under our generated tests (see Table~\ref{tab:model_quality}) are \emph{lower than} the original test accuracy (see Table~\ref{tab:model_summary_1}). On the contrast, most of the absolute accuracy differences between instances A and B under our generated tests \emph{outnumber} those under original test data.
This indicates that the generated tests can better distinguish the qualities of instances A and B, and instance C is able to generate high quality tests than the other two, which is consistent with our expectations. 

\vspace{2mm}
% \vspace{2mm}
\begin{tcolorbox}[size=title]
{\textbf{Answer to RQ 2:} {\tool} facilitates the model quality evaluation through guided fuzz testing. The tests generated with different coverage guidance exhibit different test capabilities, providing different feedback to the model quality.} %
\end{tcolorbox}

\subsection{DNN Erroneous Behavior Detection}
\label{subsec:rq3}

\begin{table}[!t]
 \vspace{2mm}
\centering

\scriptsize

\caption{The number of unique error triggering tests generated by {\tool} with different coverage guidance.}
\vspace{-2mm}
\begin{center}
\begin{tabular}{lllllll}
\hline 
\textbf{DNN} & \multicolumn{6}{c}{\textbf{Unique Error Triggering Tests (unit in 1 k)}}\tabularnewline
\cline{2-7} 
\textbf{Models} & \textbf{NC} & \textbf{KMNC} & \textbf{NBC} & \textbf{SNAC} & \textbf{TKNC} & \textbf{BKNC}\tabularnewline
\hline 
\textbf{LeNet-1} & 6.9 & 1.1 & 6.7 & 8 & 6.8 & 8.6\tabularnewline
\textbf{LeNet-4} & 4.7 & 0.6 & 2.9 & 3.3 & 2.2 & 4.5\tabularnewline
\textbf{LeNet-5} & 2.9 & 0.7 & 3.1 & 3.3 & 1.3 & 3.2\tabularnewline
\textbf{RN-20} & 0 & 7.0 & 7.8 & 7.9 & 6.1 & 7.2\tabularnewline
\textbf{VGG-16} & 2.2 & 6.3 & 8.1 & 8.5 & 6.8 & 8.3\tabularnewline
\textbf{MobileNet} & 1.5 & 10.8 & 11.6 & 9.6 & 16.6 & 13.8\tabularnewline
\textbf{RN-50} & 1.3 & 8.1 & 8.8 & 8.8 & 9.7 & 8.6\tabularnewline
\hline 
\textbf{SUM} & 19.5 & 34.6 & 49 & 49.4 & 49.5 & 54.2\tabularnewline
\hline 
\end{tabular}

\end{center}

\label{tab:erroneous}
\vspace{-4mm}
\end{table}

\textbf{To answer RQ 3}, during the fuzz testing process of {\tool}, we continuously collect the generated tests that trigger erroneous behaviors of DNNs. Since our metamorphic mutation performs constraint-based transformation on inputs, to ensure no changes of the semantic between the original image and the transformed one, we perform prediction check on images before and after transformation in batch and record the tests that trigger the erroneous behaviors of the DNN under test. The detected erroneous behaviors from proposed  coverage criteria for each model is shown in Table~\ref{tab:erroneous}.\footnote{\scriptsize Note that once error-trigger tests are generated, they are recorded for further processing without putting them back into the batch pool.} Consequently, it is not surprising that {\tool} successfully generates tests to trigger the erroneous behaviors of DNNs. The recent work~\cite{tian2018deeptest, pei2017deepxplore} have already shown that testing only based on neuron coverage already generates thousands of erroneous triggering tests.

There appears a case on RN-20 with the neuron coverage $0$. After generating 24 batches, the tests generation converges (\ie, new seeds cannot cover new coverage in terms of NC). Thus the metamorphic mutation is always run on existing batches, and new seeds are always {\it one-time} mutated from existing batches. As a result, in this case, one transformation under our conservative strategy (\cf Section~\ref{sec3:mutation}) is difficult to generate erroneous triggering tests.

\vspace{2mm}
\begin{tcolorbox}[size=title]
{\textbf{Answer to RQ 3:} {\tool} can effectively generate tests to trigger erroneous behaviors of the DNN under tests, which also scales well to practical-sized datasets and DNN models.}
\end{tcolorbox}

\subsection{Defect Detection under Controlled DNN Quantization Settings}
\label{subsec:rq4}

\textbf{To answer RQ 4}, recently there exists a strong demand to deploy DNN solutions on diverse platforms such as mobile device, edge computing device. Due to the computation and power limitation, a common practice is to quantize the DNN model from high precision floating to a lower precision form, to reduce the size for deployment. However, the quantization could introduce potential unexpected erroneous behaviors. An effective test suite should be able to capture such error cases as feedback to DL developer for further analysis and debugging. In this research question, we investigate whether {\tool} is useful to detect potential defects during quantization. 

For each of studied DNN model in Table~\ref{tab:model_summary_1}~(that is 32-bit floating point precision), we perform quantization with 3 configurations: (1) randomly sample $1\%$ of weights to truncate 32-bit floating point to 16-bit, resulting a mixed precision DNN model, (2) randomly sample $50\%$ weights to truncate 32-bit floating point to 16-bit, and (3) truncate all weights from 32-bit floating point to 16-bit.\footnote{\scriptsize Due to randomness of the first two configurations, we repeat the sampling procedure 5 times to average the results.}

Notice that the initial seeds of each dataset cannot detect the erroneous behavior before and after quantization.
Then, we reuse the tests generated by DeepHunter to evaluate quantized models, the results are summarized in Table~\ref{tab:quantization}. In all cases, {\tool} enables to detect the potential minor erroneous behaviours introduced during quantization.  

In many of the configurations, the tests generated with NBC and SNAC guidance detect more erroneous issues. One potential reason is that the tests with higher NBC and SNAC tends to cover the corner-region behavior of neurons, which could potentially trigger the erroneous behaviors of quantized model. Another interesting founding we found that the number error trigger tests for full quantization DNN model could sometimes be smaller than the mix-precision quantization counterparts. For example on LeNet-4 TKCN configuration, we found that 31 erroneous behavior on full quantization DNN model, while averaged 33 erroneous behavior on $50\%$. Intuitively, the large quantization ratio, more weights lose precision, and more erroneous behavior could be introduced. However, our evaluation results hints that sometimes the error introduced by more weight precision loss might cancel each other and obtain an less erroneous quantized version.

\vspace{2mm}
\begin{tcolorbox}[size=title]
{\textbf{Answer to RQ 4:} {\tool} can effectively detect potential defects introduced during DNN quantization, albeit a minor precision loss.} %~(\ie, $1\%$~truncation).}
\end{tcolorbox}

\begin{table}[!t]
\vspace{2mm}
\centering
\scriptsize
\caption{The number of sensitive defects are detected by {\tool} during DNN model quantization, with full quantization from 32-bit to 16-bit floating conversion, as well as with mixed precision with random parts of weights quantized. The number of defects for 1\% and 50\% quantization ratio are averaged detected defects over five runs.}
%\vspace{-3mm}
\begin{center}
\begin{tabular}{lcllllll}
\hline 
\textbf{DNN} & \multirow{1}{*}{\textbf{Quan.}} & \multicolumn{6}{c}{\textbf{Number Defects Detected}}\tabularnewline
\cline{3-8} 
\textbf{Models} & \textbf{Ratio (\%)} & \textbf{NC} & \textbf{KMNC} & \textbf{NBC} & \textbf{SNAC} & \textbf{TKNC} & \textbf{BKNC}\tabularnewline
\hline 
 & 1 & 9 & 17 & 61 & 63 & 18 & 21\tabularnewline
\textbf{LeNet-1} & 50 & 43 & 79 & 111 & 141 & 75 & 67\tabularnewline
 & 100 & 36 & 77 & 107 & 164 & 82 & 62\tabularnewline
\hline 
 & 1 & 17 & 10 & 9 & 16 & 11 & 39\tabularnewline
\textbf{LeNet-4} & 50 & 31 & 43 & 38 & 84 & 33 & 65\tabularnewline
 & 100 & 25 & 45 & 43 & 85 & 31 & 55\tabularnewline
\hline 
 & 1 & 2 & 7 & 16 & 16 & 13 & 7\tabularnewline
\textbf{LeNet-5} & 50 & 22 & 46 & 91 & 45 & 51 & 24\tabularnewline
 & 100 & 23 & 49 & 100 & 46 & 53 & 28\tabularnewline
\hline 
 & 1 & 0 & 14 & 8 & 6 & 6 & 15\tabularnewline
\textbf{RN-20} & 50 & 0 & 58 & 62 & 40 & 44 & 57\tabularnewline
 & 100 & 0 & 64 & 68 & 42 & 46 & 71\tabularnewline
\hline 
 & 1 & 1 & 5 & 7 & 7 & 11 & 9\tabularnewline
\textbf{VGG-16} & 50 & 3 & 48 & 36 & 34 & 39 & 44\tabularnewline
 & 100 & 5 & 44 & 38 & 41 & 38 & 52\tabularnewline
\hline 
 & 1 & 89 & 46 & 64 & 33 & 84 & 53\tabularnewline
\textbf{MobileNet} & 50 & 400 & 783 & 880 & 709 & 1,198 & 872\tabularnewline
 & 100 & 435 & 819 & 751 & 569 & 1,113 & 830\tabularnewline
\hline 
 & 1 & 7 & 11 & 11 & 6 & 22 & 15\tabularnewline
\textbf{RN-50} & 50 & 11,805 & 41,217 & 37,822 & 30,009 & 58,796 & 47,703\tabularnewline
 & 100 & 11,793 & 41,132 & 37,810 & 29,979 & 58,747 & 47,712\tabularnewline
\hline 
\end{tabular}
\end{center}
%\vspace{-2mm}
\label{tab:quantization}
 \vspace{-2mm}
\end{table}

\subsection{Discussion and Threats to Validity}

We perform extensive study on fuzz testing using 6 coverage criteria for guidance. 
In this section, we tend to discuss the potential effects of studied coverage criteria as feedback to guide fuzz testing, based on our experimental results.

From the coverage results (\cf Table~\ref{tab:coverage_result}) as well as the corresponding criteria definition, we find that KMNC is a fine-grained criterion, representing $k$-multisection of neurons, which easily facilitates to generate interesting tests. For example, the average coverage gain of KMNC is $47\%$, outnumbering the others. On the other hand, NC is a relatively coarse-grained criterion which represents records the ratio of activated neurons. Due to this~(see Table~\ref{tab:model_quality}), the fuzz testing guided by NC cannot generate effective results to evaluate the models with various quality. The results in Table~\ref{tab:model_summary_1} and Table~\ref{tab:controlled_model_summary} also show that NC is less effective in error triggering test detection and sensitive defect detection.

In comparison to KMNC and NC, the other four criteria show different behavior to guide fuzz testing, some of which could be difficult to cover such as SNAC and NBC.
They tend to guide fuzz testing in generating corner-case tests, so that to trigger more errornous behaviors
In Table~\ref{tab:model_summary_1}, more error-triggering tests are generated by the four than those by KMNC and NC in many cases.
Meanwhile, KMNC is a fine-grained coverage and able to generate tests that capture a large scope of major functional behaviors of DNNs. For the case of VGG16 in Table~\ref{tab:model_quality}, KMNC can more obviously distinguish accuracy of instances A and B with its generate tests compared with ones generated by the other criteria whose accuracy difference is about 1\%. 
On other models, BNC, SNAC, TKNC, and BKNC perform well in many cases. Our in-depth investigation reveals the possible reason that the tests generated from these four criteria are more likely to be the error triggering tests for instance A or B.

The selection of the subject datasets and DNN models could be a threat to validity. In this paper, we try to counter this issue with 3 well-studied datasets with diverse complexity. For DNN models on MNIST and CIFAR-10, we follow the common machine learning training practice to obtain DNN models achieved with competitive performance. On ImageNet dataset, we select the well-pretrained models from Keras (ver.2.1.3) release. 
Another threat could be the randomness of the weight sampling in the mixed precision quantization, we counter this issue by repeating the same setting five times and averaging the results.

\section{Related Work}
\label{sec:related_work}
In this section, we review the related work in the following three aspects: fuzz testing in traditional software, testing and verification of DNNs, and adversarial deep learning.
\subsection{Fuzz Testing in Traditional Software}

Fuzz testing has been widely used to safeguard software quality. Coverage guided grey-box fuzzing frameworks,  such as AFL~\cite{AFL}, libFuzzer~\cite{libFuzzer}, honggfuzz~\cite{honggfuzz}, and FOT~\cite{fse2018FOT} have been quite successful in detecting thousands of bugs in traditional software. 
On top of those, power scheduling~\cite{ccs2016AFLFast} and some other techniques, such as \cite{ccs2017AFLGo}, \cite{ccs2018Hawkeye}, have been demonstrated to be effective to guide the fuzzing procedure for different fuzzing purposes, such as increasing code coverage from low frequency execution traces or improving directedness for directed fuzzing scenarios. 

On the other hand, several other methods have been proposed to improve the mutation quality by providing structure aware mutation strategies, including LangFuzz~\cite{usenix2012langFuzz} and Skyfire~\cite{sp2017Skyfire} which generate or mutate the seeds according to some predefined grammars, or the libprotobuf mutator~\cite{libprotobuf-mutator} which could be used to mutate protobuf supported formats. Compared to dumb mutators, more meaningful mutants can be generated to pass validity checks and detect more deeper bugs. 

Other fuzzing techniques are also proposed to detect functionality bugs. For example, NEZHA~\cite{sp2017NEZHA} has been used to exploit the behavioral asymmetries between test programs to focus on inputs that are more likely to trigger logic bugs. The consistency of behaviors between different implementations serves as the oracle to detect the functionality bugs. 

Finally, some works utilize machine learning or deep learning techniques to improve the effectiveness of fuzzing~\cite{DBLP:journals/corr/GodefroidPS17,cloudfuzz,ExploitMeter,DBLP:journals/corr/abs-1711-04596,Cummins:2018:CFT:3213846.3213848,NEUZZ}. Different from these works, we attempt to perform fuzz testing on DNNs instead of leverage deep learning to fuzz traditional software.

\subsection{Testing and Verification of DL Systems}

\noindent \textbf{Testing.} 
DeepXplore~\cite{pei2017deepxplore} proposed a white-box differential testing technique to generate test inputs that potentially trigger inconsistencies between various DNNs; such inconsistencies may identify incorrect behaviors. They also investigated the usefulness of neuron coverage to measure how well the internal logic of a DNN is tested. In Tian and Pei~\etal following work DeepTest~\cite{tian2018deeptest}, they further leveraged neuron coverage to guide testing of DNN-driven autonomous cars. DeepTest adopts the domain-specific metamorphic relations between the car behaviors across different input images to detect erroneous behaviors in a single DNN model, whereas DeepXplore~\cite{pei2017deepxplore} requires to check multiple DNNs.

DeepCover~\cite{2018arXiv180304792S} adapted MC/DC test criteria~\cite{KellyJ.:2001:PTM:886632} for DNNs, they showed its usefulness on small-scale neural networks~(with no more than $500$ neurons and $5$ layers). Whether such criteria can scale to real-world sized DNN software remains to be investigated.\footnote{\scriptsize We have intended to include MC/DC criteria into {\tool}. However, such coverage analysis on the large-scale seed batch in DNNs is computationally expensive, and we leave the efficient MC/DC integration in future work.} DeepGauge~\cite{ma2018deepgauge} generalized the concept of neuron coverage and proposed a set of $5$ coverage criteria based on neuron numerical outputs. They have demonstrated that DeepGauge scales well to practical sized DNN models~(\eg, VGG-19, ResNet-50) and could capture erroneous behavior introduced by four state-of-the-art adversarial test generation techniques~(\ie, FGSM, BIM, JSAM, and CW). 
DeepMutation~\cite{ma2018deepmutation}, introduces a set of fault inject operators to generate mutant DNN models for test data quality evaluation. However, similar to traditional mutation testing, DeepMutation could be computationally intensive, since large amounts of mutant DNN models need to be generated, each of which is evaluated against the target test set. DeepCT performs combinatorial testing of DNN models to balance the huge input and latent space, and testing effectiveness~\cite{ma2018combinatorial}

Different from the existing works, this paper tends to propose a scalable and general-purpose coverage-guided fuzz testing framework for DNN software. We have integrated existing scalable coverage criteria into {\tool} in order to guide testing and defect detection in large scale.

A concurrent work, TensorFuzz \cite{odena2018tensorfuzz}, tries to debug neural networks with coverage-guided fuzzing.
{\tool} differs from TensorFuzz mainly in three aspects. %while sharing the same root. To name a few, 
In TensorFuzz, the mutator provides only one type of mutation, which is additive noise, to the inputs. In {\tool}, the mutator is entrusted with eight semantic-preserving metamorphic mutation types based on global and local image transformation, resulting in metamorphic inputs that are both diversified and plausible. 
Furthermore, in TensorFuzz, the feedback relies solely on one criterion, which is the basic neuron coverage. In {\tool}, instead, we are employing a set of six multi-granularity neuron coverage criteria for providing multi-faceted feedback to the fuzzer. Most importantly, {\tool} also differs from TensorFuzz with regards to the scope of measurement. In fact, the focus of our paper is to take a large-scale empirical study on multiple coverage to investigate their usefulness to guide test generation towards detect potential issues introduced during DNN development and deployment.

\noindent \textbf{Verification.}
The reliability of DNNs has been investigated by recent work with formal guarantees~\cite{DBLP:journals/corr/abs-1710-07859,LA10,GCDKM17,DGCC17,TMDPSM18,KYJS17}. 
Pulina \etal~\cite{LA10} proposed an abstraction-refinement approach to verify safety of a neural network with $6$ neurons; Reluplex \cite{GCDKM17} adopted an SMT-based approach on a neural network with $300$ ReLU nodes. DeepSafe \cite{DGCC17} tried to identify safe regions in the input space using Reluplex as its core. 
A more recent work AI$^{2}$ \cite{TMDPSM18} proposed an  abstract interpretation technique to verify DNN software, through a well designed abstract domains and transformation operators.
Since DNN software often handles high-dimensional input has large runtime internal states, designing more scalable and general verification methods towards complex real-world sized DNNs is challenging but important.

This paper further pushes quality assurance of DNNs from the automated fuzz testing perspective, by examining whether coverage guided fuzz testing could be a useful potential software quality assurance technique for DNN software. {\tool} could scale to ImageNet like pratical dataset with large DNN models like ResNet-50. Whether the potential integration of existing formal verification and fuzz testing is possible could be an interesting direction to explore.

\subsection{Adversarial Deep Learning} 

A plethora of research has shown that carefully-crafted adversarial examples can undermine the robustness of DNNs~\cite{goodfellow2014explaining,iclr18-ae-boundary-analysis,iclr18-a-boundary-attack,iclr18-ae-natural,iclr18-ae-spatial,szegedy2014intriguing,carlini2016towards,carlini2017adversarial}. In response to these attacks, several defenses have been proposed, such as ensemble method~\cite{iclr18-ab-ensemble}, GAN-based defense~\cite{iclr18-b-defense-gan,iclr18-b-pixeldefend}, a certified approach~\cite{iclr18-b-certified}, game theoretic based defense~\cite{iclr18-b-stochastic}, stochastic quantization method~\cite{iclr18-ab-binarized}, and so on~\cite{iclr18-b-counter,iclr18-b-thermo,iclr18-b-resistant}. However, it has been shown that none of these defenses or detection methods is robust enough against adaptive attacks~\cite{carlini2017adversarial}.

Different from adversarial techniques, {\tool} generates tests to detect both potential defects introduced during DNN development and quantization phase for deployment. In addition, these error triggering tests generated by {\tool} are not limited to adversarial tests, which makes {\tool} more general and promising than existing defenses or detection methods.

\section{Conclusion and Future Work}
\label{sec:conclusion}

Deep learning has seen tremendous success over the past decade, and has become the driving force for many novel intelligence applications. However, the quality assurance technique for DL is still at its early stage, and scalable DL testing framework is highly demanding. 
In this paper, we have proposed a coverage-guided fuzz testing framework for DNN software that systematically generates tests for detecting potential defects introduced during the DNN development and deployment phase. We have conducted a large-scale study to demonstrate its usefulness in facilitating defect detection, model quality evaluation, \etc, with data complexity increasing from MNIST to practical sized ImageNet.

Due to the computation resource limitation, we mainly focus on the investigation of how each single coverage criteria contributes to the effectiveness of {\tool}, and on the opposite side, whether {\tool} can  effectively improve the corresponding coverage. How to intelligently combine the multiple criteria to further enhance the testing performance would be our future work. Furthermore, we intend to investigate how to integrate {\tool} into DNN development and deployment practice, by providing useful feedback to help DL developers to enhance the DNN software quality. 
Since the investigation on quality assurance of deep learning is still at an early stage, we hope that {\tool} can benefit both SE and AI communities, and facilitate further extensive studies towards constructing high quality DNN software.

\bibliographystyle{IEEEtran}
\bibliography{ref}

\end{document}